\documentclass[12pt,psfig]{article}
\usepackage{times,amsmath,epsfig,cite,subfigure}
\usepackage{latexsym,amssymb}
\usepackage{xspace,amsmath,epsfig,syntonly,psfrag}
\newcommand{\lettrineo}[2]{%
\settowidth{\lwidth}{#2\kern2pt}%
\noindent\hangindent\lwidth\hangafter-#1\hskip-\lwidth%
\smash{\hbox to\lwidth{\raise7pt\vtop{\null\hbox{#2}}%
\hfill}}\ignorespaces}

\newfont{\HUGEfonto}{cmr17 scaled \magstep3}  
\DeclareSymbolFont{AMSa}{U}{msa}{m}{n}
\DeclareMathSymbol{\blacksquare}  {\mathord}{AMSa}{"04}

\newfont{\bbb}{msbm10 scaled 500}

\newfont{\bb}{msbm10 scaled 1100}
\newcommand{\CC}{\mbox{\bb C}}
\newcommand{\RR}{\mbox{\bb R}}
\newcommand{\ZZ}{\mbox{\bb Z}}

\newcommand{\mb}{\mathbf }

\newcommand{\rv}{{\bf r}}

\newcommand{\uv}{{\bf u}}
\newcommand{\wv}{{\bf w}}
\newcommand{\vv}{{\bf v}}
\newcommand{\xv}{{\bf x}}
\newcommand{\yv}{{\bf y}}
\newcommand{\zv}{{\bf z}}
\newcommand{\zerov}{{\bf 0}}

\newcommand{\Bm}{{\bf B}}

\newcommand{\Dm}{{\bf D}}

\newcommand{\Fm}{{\bf F}}
\newcommand{\Gm}{{\bf G}}
\newcommand{\Hm}{{\bf H}}
\newcommand{\Id}{{\bf I}}

\newcommand{\Pm}{{\bf P}}

\newcommand{\Cc}{{\cal C}}

\newcommand{\Nc}{{\cal N}}

\newcommand{\Rc}{{\cal R}}

\newcommand{\Uc}{{\cal U}}

\newcommand{\etav}{\hbox{\boldmath$\eta$}}
\newcommand{\lambdav}{\hbox{\boldmath$\lambda$}}

\newcommand{\diag}{{\hbox{diag}}}

\newcommand{\beq}{\begin{equation}}
\newcommand{\enq}{\end{equation}}
\newcommand{\beqa}{\begin{eqnarray}}
\newcommand{\enqa}{\end{eqnarray}}

\newcommand{\beql}[1]{\begin{equation}\label{#1}}

\newcommand{\be}{\beta}

\newcommand{\qed}{\hfill $\Box$}

\newtheorem{theorem}{Theorem}
\newtheorem{lemma}[theorem]{Lemma}
\newenvironment{proof}{{\sl Proof\/}:\ \ }{\qed\vspace{\baselineskip}}

\def\bbC{{\sf C}\kern -6pt {\sf C}}
\def\bbF{{\sf F}\kern -5pt {\sf F}}
\def\bbR{{\sf R}\kern -6pt {\sf R}}
\def\bbZ{{\sf Z}\kern -5pt {\sf Z}}
\def\sfbegin{\begingroup\sf}
\def\sfend{\endgroup}

\def\be{\begin{eqnarray*}}
\def\ee{\end{eqnarray*}}

\title{Cooperative Lattice Coding and Decoding}
\author{Arul~Murugan, Kambiz~Azarian, and Hesham~El~Gamal}
\date{}
\begin{document}
\maketitle
\begin{abstract}
A novel lattice coding framework is proposed for outage-limited
cooperative channels. This framework provides practical
implementations for the optimal cooperation protocols proposed by
Azarian~{\em et al.}. In particular, for the relay channel we
implement a variant of the dynamic decode and forward protocol,
which uses orthogonal constellations to reduce the channel seen by
the destination to a single-input single-output time-selective one,
while inheriting the same diversity-multiplexing tradeoff. This
simplification allows for building the receiver using traditional
belief propagation or tree search architectures. Our framework also
generalizes the coding scheme of Yang and Belfiore in the context of
amplify and forward cooperation. For the cooperative multiple access
channel, a tree coding approach, matched to the optimal linear
cooperation protocol of Azarain~{\em et al.}, is developed. For this
scenario, the MMSE-DFE Fano decoder is shown to enjoy an excellent
tradeoff between performance and complexity. Finally, the utility of
the proposed schemes is established via a comprehensive simulation
study.
\end{abstract}

\section{Introduction}\label{sec:intro}
Lately, cooperative communications has been the focus of intense
research activities. As a consequence, we now have a wealth of
results covering a wide range of channel models and/or system design
aspects~\cite{AES:05,SEA1,SEA2,xWT02,stefanov1,nosratinia1,duman1,MS03,HM03,YB:05}.
In this paper we focus on outage-limited
cooperative channels, where a slow fading model is assumed for both
relay and cooperative multiple access (CMA) scenarios. We further
impose the half duplex constraint limiting each node to either
transmit or receive at any point in time. The primary goal in this
setting is to construct strategies that exploit the available
\emph{cooperative diversity} with a \emph{reasonable}
encoding/decoding complexity. Our work is inspired by the
cooperation protocols of Azarian~{\em et al.} in \cite{AES:05}. In
particular, for the relay channel we implement a novel variant of
the dynamic decode and forward (DDF) strategy that, through
judicious use of orthogonal space-time constellations, reduces the
channel seen by the destination to a single-input single-output
(SISO) time-selective channel. This variant achieves the excellent
diversity-multiplexing tradeoff (DMT) of the DDF protocol
\cite{AES:05}, while minimizing the decoding complexity at both the
relay and the destination nodes. We then present a tree
coding/decoding implementation of the optimal cooperative multiple
access (CMA) strategy proposed in \cite{AES:05}. This implementation
employs the minimum mean square error decision feedback equalizer
(MMSE-DFE) Fano decoder \cite{seq}, to approximate the
maximum-likelihood (ML) performance at a much lower complexity. In
summary, our contributions in this paper are twofold. First, we
establish the practical value of the information-theoretically
optimal cooperation protocols proposed in~\cite{AES:05}. This goal
is accomplished by constructing low complexity lattice
coding/decoding implementations of the DDF and CMA cooperation
strategies, which are shown to significantly outperform the recently
proposed codes by Yang and Belfiore \cite{YB:05}. Second, we
elucidate the performance-complexity tradeoff and the parameters
controlling it, in different cooperation settings.

The rest of the paper is organized as follows. Section~\ref{sec:sys}
introduces our system model and notations. In
Section~\ref{sec:lattice}, we briefly review the lattice
coding/decoding framework adopted in our work. The proposed
coding/decoding schemes for the outage-limited relay channel are
detailed in Section~\ref{sec:relay}. Section~\ref{sec:cma} is
devoted to the description of our tree coding/decoding approach for
the CMA channel. Finally, we offer some concluding remarks in
Section~\ref{sec:conc}.

\section{System Model and Notation}\label{sec:sys}
In this section, we state the assumptions that apply to the two
scenarios considered in this paper (i.e., relay and CMA channels).
Assumptions pertaining to a specific channel will be given in the
related section. All channels are assumed to be flat Rayleigh-fading
and quasi-static, i.e., the channel gains remain constant during one
codeword and change independently from one codeword to the next.
Furthermore, the channel gains are mutually independent with unit
variance. The additive noises at different nodes are zero-mean,
mutually-independent, circularly-symmetric, and white
complex-Gaussian. The variances of these noises are proportional to
one another such that there are always \emph{fixed} offsets between
the different signal-to-noise ratios (SNRs). All nodes have the same
power constraint, have a single antenna, and operate synchronously.
Only the receiving node of any link knows the channel gain; no
feedback to the transmitting node is permitted. All cooperating
partners operate in the half-duplex mode, i.e., at any point in
time, a node can either transmit or receive, but not both. This
constraint is motivated by, e.g., the typically large difference
between the incoming and outgoing signal power levels. Next we
summarize the notation used throughout the paper.
\begin{enumerate}
\item The SNR of a link, $\rho$, is defined as
\begin{align}
\rho &\triangleq \frac{E}{\sigma^2}, \label{eq:44}
\end{align}
where $E$ denotes the average energy available for transmission of
a symbol across the link and $\sigma^2$ denotes the variance of
the noise observed at the receiving end of the link. We say that
$f(\rho)$ is \emph{exponentially equal to} $\rho^b$, denoted by
$f(\rho) \dot{=} \rho^b$, when
\begin{align}
\lim_{\rho \rightarrow \infty} \frac{\log (f(\rho))}{\log (\rho)}
&= b. \label{eq:45}
\end{align}
In \eqref{eq:45}, $b$ is called the \emph{exponential order} of
$f(\rho)$. $\dot{\leq}$ and $\dot{\geq}$ are defined similarly.

\item Assuming that $g$ is a complex Gaussian random variable with zero
mean and unit variance, the exponential order of $1/|g|^2$ is
defined as
\begin{align}
v &= -\lim_{\rho \rightarrow \infty}
\frac{\log(|g|^2)}{\log(\rho)}. \label{eq:46}
\end{align}
The probability density function (PDF) of $v$ has the following
property~\cite{ZT:02}
\begin{align}
p_{v} &\dot{=}
\begin{cases}
\rho^{-\infty}= 0, &\text{~for~} v<0, \\
\rho^{-v}, &\text{~for~} v\geq 0
\end{cases}. \label{eq:47}
\end{align}

\item Consider a family of codes $\{C_{\rho}\}$ indexed by
operating SNR $\rho$, such that the code $C_{\rho}$ has a rate of
$R(\rho)$ bits per channel use (BPCU) and ML error probability
$P_e(\rho)$. For this family, the multiplexing gain $r$ and the
diversity gain $d$ are defined as
\begin{align}
&r \triangleq \lim_{\rho\rightarrow\infty}\frac{R(\rho)}{\log\rho}
, &d \triangleq
-\lim_{\rho\rightarrow\infty}\frac{\log(P_e(\rho))}{\log\rho} .
\label{eq:1}
\end{align}

\item The problem of characterizing the optimal DMT in a
point-to-point communication system over a coherent quasi-static
flat Rayleigh-fading channel was posed and solved by Zheng and Tse
in \cite{ZT02}. For a MIMO communication system with $M$ transmit
and $N$ receive antennas, they showed that, for any $r \leq
\min\{M,N\}$, the optimal diversity gain $d^*(r)$ is given by the
piecewise linear function joining the $(r,d)$ pairs $(k,(M-k)(N-k))$
for $k=0,...,\min\{M,N\}$, provided that the code-length $l$
satisfies $l\ge M+N-1$.

\item We say that protocol $A$ \emph{uniformly dominates} protocol
$B$ if, for any multiplexing gain $r$, $d_A(r) \geq d_B(r)$.

\item We say that protocol $A$ is \emph{Pareto optimal}, if there
is no protocol $B$ that dominates protocol $A$ in the Pareto sense.
Protocol $B$ is said to dominate protocol $A$ in the Pareto sense if
there is some $r_0$ for which $d_B(r_0) > d_A(r_0)$, but no $r$ such
that $d_B(r) < d_A(r)$.

\item We denote the ratio of the destination variance to that of
  inter-user noise variance by $c$, {\it i.e.,} $c=\sigma_v^2/\sigma_w^2$.

\item Throughout the sequel, vectors are denoted by bold lowercase
characters (e.g., $\xv$), and matrices are denoted by bold uppercase
characters (e.g., $\Hm$). $\ZZ$, $\RR$, ${\mathbb C}$ refer to the
ring of integers, field of real numbers, and field of complex
numbers, respectively. We refer to the identity matrix of dimension
$M$ as $\Id_{M}$ and Kronecker product by $\otimes$. We also use
$(x)^+$ to mean $\max\{x,0\}$, $(x)^-$ to mean $\min\{x,0\}$ and
$\lceil x \rceil$ to mean nearest integer to $x$ towards plus
infinity.

\end{enumerate}

\section{Lattice Coding and Decoding}\label{sec:lattice}
An $m$-dimensional lattice $\Lambda \subset\RR^m$ is the set of
points
\begin{equation} \label{lattice}
\Lambda = \{ \lambdav = \Gm \uv \; : \; \uv \in \ZZ^m \}
\end{equation}
where $\Gm \in \RR^{m\times m}$ is the lattice generator matrix.
Let $\etav \in \RR^m$ be a vector and $\Rc$ a measurable region in
$\RR^m$ then a lattice code $\Cc(\Lambda,\eta,\Rc)$ is defined as
the set of points of the lattice translate $\Lambda + \etav$ inside
the {\em shaping
  region} $\Rc$~\cite{forney:cosetI}, i.e.,
\begin{equation} \label{code}
\Cc(\Lambda,\etav,\Rc) = \{\Lambda + \etav\} \cap \Rc.
\end{equation}
Here, we focus on construction ``A'' lattice
codes~\cite{loeliger}. In this construction, $\Lambda = C +
Q\ZZ^m$ with $C \subseteq \ZZ_Q^m$ being a linear code over
$\ZZ_Q$ and $Q$ is a prime. The generator matrix of $\Lambda$ is
given by
\begin{equation} \label{gmatrix}
\Gm = \left [ \begin{array}{cc}
\Id & \zerov \\
\Pm & Q \Id \end{array} \right ],
\end{equation} where $[\Id, \Pm^{\rm T}]^{\rm T}$ is
the generator matrix of $C$ (in a systematic
form)~\cite{loeliger}. The shaping region $\Rc$ of the lattice
code $\Cc(\Lambda,\etav,\Rc)$ can be 1) the $m$-dimensional sphere,
2) the fundamental Voronoi region of a sublattice $\Lambda'
\subset \Lambda$, or 3) the $m$-dimensional hypercube. These three
alternatives provide a tradeoff between performance and encoding
complexity. The spherical shaping approach yields the maximum
shaping gain but suffers from the largest encoding complexity.
Encoding is typically done via look-up tables which limits this
approach to short block lengths (since the number of entries in
the table grows exponentially with the block length). Voronoi
shaping offers a nice compromise between performance and
complexity. In this approach, the encoding complexity is
equivalent to that of $m$-dimensional vector quantization which,
loosely speaking, ranges from polynomial to linear (in the block
length) based on the choice of the shaping lattice. On the other
hand, with large enough $m$ and the appropriate choice of
$\Lambda'$, the shaping gain approaches that of the sphere. The
third alternative, i.e., hypercubic shaping, allows for the lowest
encoding complexity, however, at the expense of a performance loss
(i.e., the worst shaping gain). As a side benefit, hypercubic
shaping also minimizes the peak-to-average power ratio. The final
ingredient in lattice coding is the translate vector $\etav$ which
is used to maximize the number of lattice points inside $\Rc$
and/or randomize the distribution of the codebook over $\Rc$
\cite{forney:cosetI,EZ03}. In fact, Voronoi coding with $\etav$
uniformally distributed over the Voronoi cell of $\Lambda'$
corresponds to the mod-$\Lambda$ approach of Erez and
Zamir~\cite{EZ03}. The proposed coding approaches in our work can
be coupled with any shaping technique and any choice for the
translate $\etav$. The optimization of these parameters is beyond
the scope of this paper, and hence, will not be considered
further. For simplicity of presentation, and implementation, we
will focus in our simulation study on hypercubic shaping.

The primary appeal of lattice codes, in our framework, stems from
their amenability to a low complexity decoding architecture, as
argued in the sequel. For decoding purposes, we express
$\Cc(\Lambda,\etav,\Rc)$ as the set of points $\xv$ given by
\begin{equation} \label{code1}
\xv = \Gm \uv+\etav, \;\;\; \mbox{for} \;\; \uv \in \Uc
\end{equation}
where $\Uc \subset \ZZ^m$ is the code {\em information set}.
Assuming that the code is used over a linear Gaussian channel,
then the input-output relation is given by
\begin{equation} \label{sys0}
\mathbf{r = H x + z}
\end{equation}
where $\rv \in \RR^n$ denotes the received signal vector, $\zv
\sim \Nc(\zerov,\sigma^2\Id)$ is the AWGN vector, and $\Hm \in {\mathbb
R}^{n\times m}$ is a matrix that defines the channel linear
mapping. In the coherent paradigm, where $\Hm$ is known to the
receiver, the maximum likelihood (ML) decoding rule reduces to
\begin{equation}
\label{sys0a} \hat{\mb u} = \textrm{arg} \min_{{\mb u}\in
{\mathcal U}} |{\mathbf r}-{\Hm}{\etav}-{\mb H}\Gm {\mb u}|^2
\end{equation}
The optimization problem in (\ref{sys0a}) can be viewed as a {\em
  constrained} version of the closest lattice point search (CLPS) with lattice
generator matrix given by ${\mathbf H}\Gm$ and constraint set
$\Uc$~\cite{seq}. This observation inspired the class of {\em
sphere decoding} algorithms (e.g.,~\cite{AEVZ02}). More recently,
a unified tree search decoding framework which encompasses the
sphere decoders as special cases was proposed in \cite{seq}. Of
particular interest to our work here is the (MMSE-DFE) Fano
decoder discovered in~\cite{seq}. In this decoder, we first {\em
preprocess} the channel matrix via the feedforward filter of the
MMSE-DFE then we apply the celebrated Fano tree search algorithm
to identify the closest lattice point. In particular, we attempt
to approximate the optimal solution for

\begin{equation}
\label{sys0b} \hat{\mb u} = \textrm{arg} \min_{{\mb u}\in \bbZ^m}
|\Fm\left({\mathbf r}-{\Hm}{\etav}-{\mb H}\Gm {\mb u}\right)|^2,
\end{equation}

where $\Fm$ is the feedforward filter of the MMSE-DFE. It is
important to note the expanded search space in (\ref{sys0b}),
i.e., we replaced ${\mathcal U}$ in (\ref{sys0a}) with $\bbZ^m$.
Relaxing this constraint results in a significant complexity
reduction since enforcing the boundary control ${\mb u}\in
{\mathcal U}$ can be computationally intensive for non-trivial
lattice codes~\cite{seq}. While the search space expansion is
another source for sub-optimality, it was shown in~\cite{last,seq}
that the loss in performance is very marginal {\bf only}when the
MMSE-DFE preprocessing is employed (i.e., with $\Fm=\Id$ one would
see a significant performance loss).

One of our main contributions is showing that the MMSE-DFE Fano
decoder yields an excellent performance-vs-complexity tradeoff when
appropriately used in cooperative channels. For more details about
this decoder, the reader is referred to~\cite{seq}.

\section{The Relay Channel}\label{sec:relay}
For exposition purposes, we limit our discussion to the single
relay scenario. The proposed techniques, however, extend naturally
to channels with an arbitrary number of relays.
\subsection{Amplify and Forward (AF) Cooperation} In \cite{AES:05}, the
non-orthogonal amplify and forward (NAF) strategy was shown to
achieve the optimal DMT within the class of AF protocols. In NAF
relaying, the source transmits on every symbol-interval in a
cooperation frame, where a cooperation frame is defined as two
consecutive symbol-intervals. The relay, on the other hand,
transmits only once per cooperation frame; it simply repeats the
(noisy) signal it observed during the previous symbol-interval. It
is clear that this design is dictated by the half-duplex
constraint, which implies that the relay can repeat at most once
per cooperation frame. We denote the repetition gain by $b$ and,
for frame $k$, the information symbols are denoted by
$\{x_{j,k}\}_{j=1}^{2}$. The signals received by the destination
during frame $k$ are given by:
\begin{align}
y_{1,k}&=g_1x_{1,k}+v_{1,k}, \label{eq:30} \\
y_{2,k}&=g_1x_{2,k}+g_2b(hx_{1,k}+w_{1,k})+v_{2,k}. \label{eq:31}
\end{align}
Note that, in order to decode the message, the destination needs
to know the relay repetition gain $b$, the source-relay channel
gain $h$, the source-destination channel gain $g_1$, and the
relay-destination channel gain $g_2$. The following result
from~\cite{AES:05} states the DMT achieved by this protocol.
\begin{theorem}
The diversity-multiplexing tradeoff achieved by the NAF relay
protocol is
\begin{align}
d(r) &= 1-r +(1-2r)^+. \label{eq:32}
\end{align}
\end{theorem}

In \cite{AES:05}, the achievability of (\ref{eq:32}) was
established using a long Gaussian codebook which spans infinitely
many cooperation frames (with the same channel coefficients). More
recently, Yang and Belfiore have proposed a novel scheme that
achieves the tradeoff in (\ref{eq:32}) by only coding over one
cooperation frame. Yang and Belfiore design is inspired by the
fact that the input-output relationship in (\ref{eq:30}) and
(\ref{eq:31}) corresponds to the following $2\times 2$ MIMO
channel\beq
\label{naf_relay_sys} \yv=\left[\begin{array}{cc} g_1 & 0 \\
\sqrt{\frac{c}{|g_2 b|^2+c}}g_2 bh &
\sqrt{\frac{c}{|g_2 b|^2+c}} g_1 \end{array}\right]\xv+\zv
\\
\enq where $\zv\in \CC^2$ is the noise vector with circularly
symmetric i.i.d. Gaussian components, $z_i\sim{\mathcal
N}_C(0,\sigma_v^2)$. One can then use any of the $2\times 2$ linear
dispersion (LD) constellations~\cite{HH00} as a cooperation scheme
in this setup. A $2\times 2$ LD constellation is obtained by
multiplying a $4$-dimensional QAM vector $\uv$ by a generator
matrix. As shown in~\cite{YB:05}, by setting the generator matrix
to that
 of the so called Golden constellation, i.e.,
\be \label{golden}
\Gm_{gc}=\frac{1}{\sqrt{5}}\left[\begin{array}{cccc} \alpha & \alpha\theta
    & 0 & 0 \\ 0 & 0 & i\bar{\alpha} & i\bar{\alpha}\bar{\theta} \\ 0 & 0 &
    \alpha & \alpha\theta \\ \bar{\alpha} & \bar{\alpha}\bar{\theta} & 0 & 0
    \end{array}\right]
\ee
where $\theta=\frac{1+\sqrt{5}}{2},
\bar{\theta}=1-\theta,\alpha=1+i\bar{\theta}$ and $\bar{\alpha}=1+i\theta$,
one can use the non-vanishing determinant property to establish the
achievability of (\ref{eq:32}) by this scheme. The lattice
decoding framework, adopted here, can be applied to this approach
by applying the appropriate scaling factor and separating the real
and imaginary parts of the received signal. This reduces
(\ref{eq:32}) to our model in (\ref{sys0}) where ${\cal
U}=\ZZ_Q^8$ corresponds to an input $Q^2$-QAM constellation.

In applications where the code word is allowed to span multiple
cooperation frames, Yang and Belfiore Golden constellation fails
to exploit the long block length in improving the coding gain. By
appealing to the lattice coding framework, one can improve the
coding gain while still using the same decoder. For example, we
can concatenate the inner Golden constellation with an outer trellis code
whose constraint length is allowed to increase with the block
length. The improved coding gain of this approach translated into
enhanced frame error rates (as validated by the numerical results
in Section~\ref{simresults}). In our design, we generate our
trellis code as a systematic convolutional code (CC) over $\ZZ_Q$.
Assuming that one code word spans $N$ cooperation frames, the
received vector can now be expressed as \beqa
\label{nafrelay_gcpluscc}
\yv&=&\Hm\Gm^\prime_{gc}\xv+\zv \\
   &=&\Hm\Gm^\prime_{gc}(\Gm_{cc}\uv+\etav)+\zv\label{lattice-cc}
\enqa where $\xv\in \RR^{2N}$ is the output of the CC, and
$(\Hm\Gm_{gc})$ is the effective channel seen by the CC. This
effective channel is obtained as follows: $\Hm$ is the real
representation of the $2\times 2$ MIMO channel in
(\ref{naf_relay_sys}), and $\Gm^\prime_{gc} =\Id_{N/2} \otimes
{\Gm}^{(r)}_{gc}$ where ${\Gm}^{(r)}_{gc}\in \RR^{8\times 8}$ is
real representation of the Golden constellation generator matrix
in (\ref{golden}). The model is (\ref{lattice-cc}) is based on
observing that the CC can be viewed as a construction A lattice
code with hypercubic shaping and a generator matrix $\Gm_{cc}$.
Now, we multiply by the feedforward filter $\Fm$ of the MMSE-DFE
for the effective channel $(\Hm\Gm_{gc})$ and then use Fano search
algorithm on the composite channel-code generator matrix
$\Fm\Hm\Gm_{gc}\Gm_{cc}$.

\subsection{Decode and Forward (DF) Cooperation}\label{df}
To the best of our knowledge, within the class of DF strategies,
the dynamic decode and forward (DDF) achieves the best
DMT~\cite{AES:05} (the same protocol was independently discovered
in different contexts~\cite{katz_shamai,MOT05}). This motivates
developing low complexity variants of this strategy which are
particularly suited for implementation. We take a step by step
approach where a number of lemmas, that characterize the
modifications needed for complexity reduction while maintaining a
good performance, are derived. For the sake of completeness, we
first describe the DDF protocol.

In the DDF protocol the source transmits data, at a rate of $R$
bits per channel use (BPCU), during every symbol-interval in the
codeword. A codeword is defined as $M$ consecutive sub-blocks,
during which all channel gains remain fixed. Each sub-block is
composed of $T$ symbol-intervals. The relay listens to the source
for enough sub-blocks until the mutual information between its
received signal and source signal exceeds $NTR$. It then decodes
and re-encodes the message using an independent Gaussian code-book
and transmits it during the rest of the codeword. We denote the
signals transmitted by the source and relay by
$\{x_k\}_{k=1}^{MT}$ and $\{\tilde{x}_k\}_{k=M'T+1}^{MT}$,
respectively, where $N'$ is the number of sub-blocks that the
relay waits before starting transmission. Using this notation, the
received signals (at the destination) can be written as
\begin{align}
y_k &=\left\{
\begin{array}{lll}
g_1x_k+v_k  &\text{for} &N'T \geq k \geq 1\\
g_1x_k+g_2\tilde{x}_k+v_k  &\text{for} &NT \geq k > M'T
\end{array} \nonumber
\right.,
\end{align}
where $g_1$ and $g_2$ denote the source-destination and
relay-destination channel gains, respectively. It is now clear
that the number of sub-blocks that the relay listens, should be
chosen according to
\begin{align}
M'&=\min\left\{M,\left\lceil\frac{MR}{\log_2{(1+|h|^2c\rho)}}\right\rceil\right\},\label{rule1}
\end{align}
where $h$ is the source-relay channel gain. In this expression,
$c=\sigma_v^2/\sigma_w^2$ denotes the ratio of the destination
noise variance, to that of the relay. The following result
from~\cite{AES:05}, describes the DMT achievable by the
DDF protocol as $T\rightarrow \infty$ and $M\rightarrow \infty$.
\begin{theorem} \label{thrm:1}
The diversity-multiplexing tradeoff achieved by the DDF protocol is
given by
\begin{align}
d(r) &= \left\{
\begin{array}{lll}
2(1-r) & \text{if} & \frac{1}{2} \geq r \geq 0 \\
(1-r)/r & \text{if} & 1 \geq r \geq \frac{1}{2}
\end{array} \label{eq:10}
\right. .
\end{align}
\end{theorem}

It is evident from the protocol description that the achievability
result in~\cite{AES:05} relied on using independent Gaussian
codebooks at the source and relay nodes. This approach will
potentially require a computationally intensive algorithm at the
destination to implement {\em joint} decoding for the source and
relay signals. Allowing the relay node to start transmission at
the beginning of every sub-block, based on the value of the
instantaneous mutual information, is another potential source for
complexity. In practice, this requires the source to use a very
high-dimensional constellation (with a very low rate code) such
that the information stream is uniquely decodable from one
sub-block if the source-relay channel is good enough. This feature
also impacts the amount of overhead in the relay-destination
packet since the destination must be information with the starting
time of relay transmission. Now, we introduce two simplifications
of the original DDF protocol that aim to lower the complexities
associated with these two properties.

\begin{enumerate}
\item After successfully decoding, the relay can correctly
anticipate the future transmissions from the source (i.e., $x_k$
for $M'T+1\leq k\leq MT$) since it knows the source codebook.
Based on this knowledge, the relay implements the following
scheme, i.e.
\begin{align}
\tilde{x}_k &= \left\{
\begin{array}{rll}
x^*_{k+1} & \text{for} & k=M'T+1,M'T+3,\cdots \\
-x^*_{k-1} & \text{for} & k=M'T+2,M'T+4,\cdots
\end{array} \label{eq:2}
\right., \text{~~and}
\end{align}
which reduces the signal seen by the destination for $M'T+1\leq
k\leq MT$ to an Alamouti constellation. \item We allow the relay
to transmit only after the codeword is halfway through,i.e., we
replace the rule in (\ref{rule1}) with
\begin{align}
M' &= \min \left\{M, \max \left\{\frac{M}{2}, \left\lceil
\frac{MR}{\log_2{(1+|h|^2c\rho)}} \right\rceil \right\}
\right\},\label{rule2}
\end{align}
\end{enumerate}

Fortunately, these modifications do not entail any loss in
performance (at least from the DMT perspective) as formalized in
the following lemma.
\begin{lemma} \label{thrm:2}
The modified (lower complexity) DDF protocol still achieves the
same DMT in Theorem~\ref{thrm:1}.
\end{lemma}
\begin{proof}
To prove the first part of the lemma, let us denote the signals
received at the destination by $\{y_k\}_{k=1}^{MT}$. Then
\begin{align}
y_k &= \left\{
\begin{array}{rll}
g_1 x_k + v_k & \text{for} & k=1,\cdots,M'T\\
g_1 x_k + g_2 x_{k+1}^* + v_k & \text{for} & k=M'T+1,M'T+3,\cdots\\
g_1 x_k - g_2 x_{k-1}^* + v_k & \text{for} & k=M'T+2,M'T+4,\cdots\\
\end{array} \nonumber
\right.
\end{align}
Now, through linear processing of $\{y_k\}_{k=1}^MT$, the
destination derives $\{\tilde{y_k}\}_{k=1}^MT$ such that
\begin{align}
\tilde{y}_k &= \left\{
\begin{array}{rll}
g_1 x_k + \tilde{v}_k & \text{for} & k=1,\cdots,M'T\\
\sqrt{|g_1|^2+|g_2|^2} x_k + \tilde{v}_k & \text{for} &
k=M'T+1,\cdots,MT
\end{array} \label{eq:3}
\right.,
\end{align}
with $\tilde{v_k}$ being statistically identical to $v_k$. Using
\eqref{eq:3}, it is straightforward to see that destination pairwise
error probability, averaged over the ensemble of Gaussian codes
(used by the source) and conditioned on a certain channel
realization, is given by
\begin{align}
P_{PE|g_1,g_2,h} &\leq \big(1+\frac{1}{2}|g_1|^2\rho \big)^{-M'T}
\big( 1+\frac{1}{2}(|g_1|^2+|g_2|^2)\rho
\big)^{-(M-M')T}.\nonumber
\end{align}
This last expression, though, is identical to the one
corresponding to the original DDF protocol (refer to
\cite{AES:05}). This means that this implementation of the DDF
strategy will achieve the same DMT as Theorem~\ref{thrm:1} as
$T\rightarrow\infty$ and $M\rightarrow\infty$ which completes the
proof of the first part.

To prove the second part, we notice that the effect of
constraining the relay to start transmission only after the
codeword is halfway through, i.e., adopting the rule
(\ref{rule2}), is to replace equation $(52)$ in \cite{AES:05} with
\begin{align}
\inf_{O^+,f} u &= \left\{
\begin{array}{lll}
0 & \text{if} & f=\frac{1}{2}\\
(1-\frac{r}{f})^+ & \text{if} & 1 \geq f > \frac{1}{2}
\end{array} \label{eq:4}
\right.,
\end{align}
where $u$ denotes the exponential order of $1/|h|^2$. Now, since
\eqref{eq:4} is different from $(52)$ in \cite{AES:05} only for
$f=1/2$, for which $\inf (v_1+v_2)$ is already equal to the
optimal value $2(1-r)$, we conclude that this restriction does not
affect the DMT achieved by the protocol.
\end{proof}

As shown in \eqref{eq:3}, the channel seen by the destination in
the modified DDF protocol is a time-selective SISO which
facilitates leveraging standard SISO decoding architectures (e.g.,
belief propagation, Fano decoder) at the destination. In addition,
by restricting the relay to transmit only after $M'\geq M/2$ means
that the constellation size can chosen such that the information
stream is uniquely decodable only after $M'=M/2$.

The next result investigates the effect of allowing the relay to
transmit only at a finite number of instants. These instants
partition the code word into $N+1$ segments which are not
necessarily equal in length.  We assume that the $j$-th segment
starts at the beginning of sub-block $N_j+1$, denote the set of
fractions  $\{f_j\}_{j=1}^N$ by to $N$ waiting fractions $f_j$
such that $f_j \triangleq \frac{N_j}{N}$ with $f_0 \triangleq 0$
and $f_{N+1} \triangleq 1$. Thus
\begin{align}
f_0 \triangleq 0 < f_1 < \cdots <f_N < f_{N+1}\triangleq 1.
\nonumber
\end{align}
The question now is how to choose $\{f_j\}_{j=1}^N$, for a finite
$N$, such that the protocol achieves the \emph{optimal} DMT. The
following lemma shows that this problem does not have a unique
optimal solution and characterizes a Pareto optimal set of
fractions.
\begin{lemma}\label{thrm:3}
For the DDF protocol with a finite $N$,
\begin{enumerate}
\item there exists no \emph{uniformly} dominant set of fractions
$\{f^u_j\}_{j=1}^N$.

\item let $f^p_1 = \frac{1}{2}$  and
\begin{align}
f^p_j &= \frac{1-f^p_{j-1}}{2-(1+\frac{1}{f^p_N})f^p_{j-1}},
\text{~~for~~} N \geq j > 1 \label{eq:5}
\end{align}
then the set of fractions $\{f^p_j\}_{j=1}^N$ is Pareto optimal,
with
\begin{align}
d^p(r) &= 1-r + (1-\frac{r}{f^p_N})^+. \label{eq:13}
\end{align}
\end{enumerate}
\end{lemma}
\begin{proof}
We note that the outage set for the DDF protocol with a general
set of waiting fractions, $\{f_j\}_{j=1}^N$, is still given by
equation $(49)$ in \cite{AES:05}, i.e.
\begin{align}
O^+ &= \{(v_1,v_2,u) \in \RR^{3+}| f(1-v_1)^+ +
(1-f)(1-\min\{v_1,v_2\})^+ \leq r \}. \nonumber
\end{align}
The only difference is that $f$ is now given by (compare with $(52)$
of \cite{AES:05})
\begin{align}
f &= \left\{
\begin{array}{lll}
f_1 & \text{if} & 1-\frac{r}{f_1} > u \geq 0\\
f_j & \text{if} & 1-\frac{r}{f_j} > u \geq (1-\frac{r}{f_{j-1}})^+\\
1   & \text{if} &                   u \geq (1-\frac{r}{f_N})^+
\end{array} \label{eq:6}
\right. .
\end{align}
Next, we split $O^+$ such that
\begin{align}
O^+ &= \cup_{j=1}^{N+1} O_j^+, \text{~~where~~} O_j^+ =
\{(v_1,v_2,u) \in O^+| f=f_j \}. \label{eq:7}
\end{align}
Now, from \eqref{eq:6} and \eqref{eq:7} we get
\begin{align}
\inf_{O_j^+} u &= \left\{
\begin{array}{lll}
0 & \text{for} & j=1\\
(1-\frac{r}{f_{j-1}})^+ & \text{for} & N+1 \geq j > 2
\end{array} \nonumber
\right. .
\end{align}
Also, since $f_j \geq \frac{1}{2}$, we have $\inf_{(v_1,v_2) \in
O_j^+} (v_1 + v_2) =(1-r)/f_j$ (refer to $(55)$ in \cite{AES:05}).
Thus
\begin{align}
d_j(r) &\triangleq \inf_{O_j^+} (v_1+v_2+u), \nonumber \\
d_j(r) &= \frac{1-r}{f_j}+(1-\frac{r}{f_{j-1}})^+. \label{eq:8}
\end{align}
But, \eqref{eq:7} along with \eqref{eq:8} results in
\begin{align}
d(r) &= \min_{f_j \geq r} d_j(r), \nonumber \\
d(r) &= \min_{f_j \geq r} \frac{1-r}{f_j}+(1-\frac{r}{f_{j-1}})^+.
\label{eq:9}
\end{align}
Now let us assume that the set of waiting fractions
$\{f^u_j\}_{j=1}^N$ is uniformly optimal. Pick $\{f_j\}_{j=1}^N$
such that $f^u_N < f_N < 1$. Then from \eqref{eq:9} we conclude that
for any $f^u_N < r < f_N$, $d^u(r)=1-r$ and $d(r)=1-r+1-r/f_N$. Thus
$d^u(r) < d(r)$, which is in contradiction with the uniform
optimality assumption of $\{f^u_j\}_{j=1}^N$. To prove the second
part of the lemma, we observe that for $N \geq j \geq 1$,
$\{f^p_j\}_{j=1}^N$ as given by \eqref{eq:5} results in
\begin{align}
d^p_{N+1}(r) &< d^p_j (r), \text{~~for~~} f_j \geq r \geq 0, r \neq
f_{j-1} \label{eq:11}
\end{align}
and
\begin{align}
d^p_{N+1}(f_{j-1}) &= d^p_j (f_{j-1}),\text{~~or}\nonumber \\
d^p_{N+1}(f_{j-1}) &=\frac{1-f_{j-1}}{f_j}. \label{eq:12}
\end{align}
Now \eqref{eq:11}, along with \eqref{eq:9} and \eqref{eq:8} proves
that $\{f^p_j\}_{j=1}^N$ achieves \eqref{eq:13}. The only thing left
is to show that $d^p(r)$ is indeed Pareto optimal, i.e., no other
set $\{f_j\}_{j=1}^N$ dominates $\{f^p_j\}_{j=1}^N$, in the Pareto
sense. To do so, we assume such a set exists and observe that since
$f_0=f^p_0=0$ and $f_{N+1}=f^p_{N+1}=1$, there should be $N+1 \geq i
\geq 1$ and $N+1 \geq \ell \geq 1$ such that
\begin{align}
&f_{i-1} \leq f^p_{\ell-1} < f^p_{\ell} < f_i, \text{~~or}
\label{eq:14} \\
&f_{i-1} < f^p_{\ell-1} < f^p_{\ell} \leq f_i. \label{eq:15}
\end{align}
Now, if \eqref{eq:14} is true, then we observe from \eqref{eq:9}
that
\begin{align}
d(f^p_{\ell-1}) \leq d_i(f^p_{\ell-1}) = \frac{1-f^p_{\ell-1}}{f_i}
< \frac{1-f^p_{\ell-1}}{f^p_{\ell}} = d^p(f^p_{\ell-1}), \nonumber
\end{align}
where we have used \eqref{eq:12} in deriving the last step. This,
however is in contradiction with the Pareto dominance of $\{f_j\}$,
since
\begin{align}
d(f^p_{\ell-1}) &< d^p(f^p_{\ell-1}). \nonumber
\end{align}
On the other hand, if \eqref{eq:15} is true, then
\begin{align}
d(f_{i-1}) \leq d_i(f_{i-1}) = \frac{1-f_{i-1}}{f_i} <
2-(1+\frac{1}{f^p_N})f_{i-1} = d^p(f_{i-1}), \nonumber
\end{align}
or
\begin{align}
d(f_{i-1}) &< d^p(f_{i-1}), \nonumber
\end{align}
which again is in contradiction with Pareto dominance of $\{f_j\}$.
This completes the proof of the second part.
\end{proof}

\begin{figure}   
\begin{center}
\epsfig{file=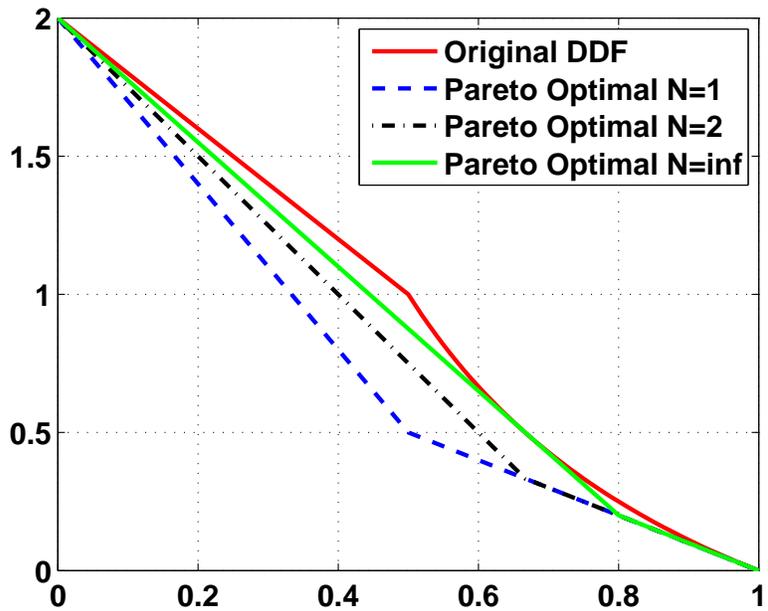,width=12.0cm}
\end{center}
\caption{Pareto optimal diversity-multiplexing tradeoff for the DDF
protocol with $N=1,2$ and $\infty$.}
\label{pareto}
\end{figure}

Figure~\ref{pareto} shows the DMT for Pareto optimal DDF protocols
with $N=1 (\{\frac{1}{2}\})$, $N=2 (\{\frac{1}{2}, \frac{2}{3}\})$
and $N=\infty$.

\subsection{Numerical Results} \label{simresults}
Throughout this section, we consider construction-A lattice codes
obtained from systematic convolutional codes (CC) over $\ZZ_Q$
with $Q$ a prime number. Unless otherwise mentioned, we choose the
SNR level of the inter-user channel to be $3$~dB higher than the
SNR at the destination. In all scenarios, the MMSE-DFE Fano
decoder uses a bias $b=1.2$ and a step-size
$\Delta=5$~\cite{MEDC05}.

\begin{figure}   
\begin{center}
\epsfig{file=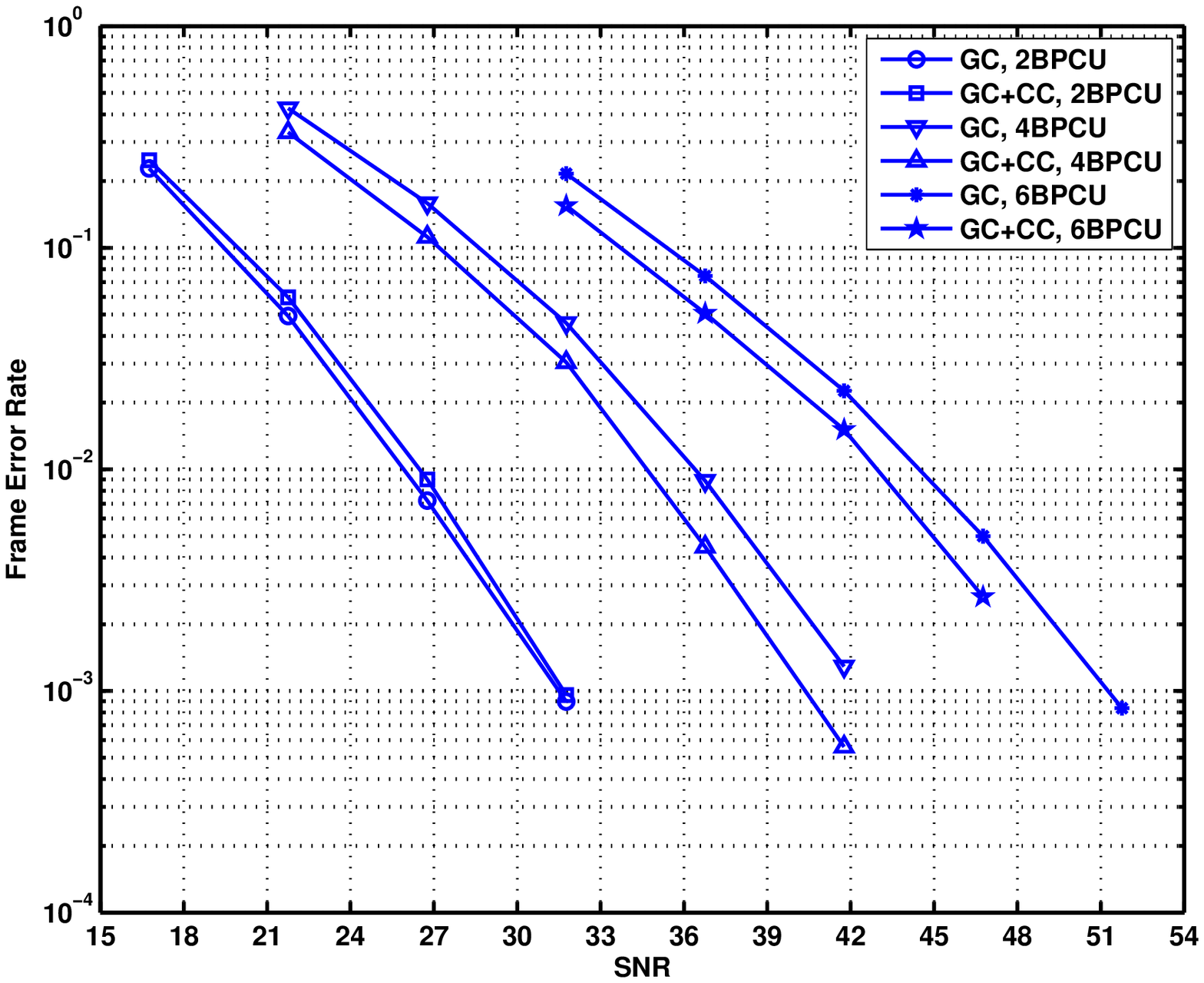,width=12.0cm}
\end{center}
\caption{Performance of NAF relay with Golden constellation, with the CC
  outer code (denoted by GC+CC) and without any outer code(denoted by GC).}
\label{nafrelay_eqpower}
\end{figure}

First, we demonstrate the performance improvement obtained by
augmenting the Golden constellation of \cite{YB:05} with a lattice
code obtained from a rate $1/2$ systematic CC. In both schemes,
the frame length is $128$. We assume that the source and the relay
transmit with equal power, and that the source transmits with same
power at all instants. Figure~\ref{nafrelay_eqpower} shows the
frame error rate (FER) of the two coding schemes as a function of
the average received SNR at the destination. For the Golden constellation,
$4-$QAM, $16-$QAM and $64-$QAM modulations for the information
symbols lead to transmission rates of $2,4$ and $6$ BPCU,
respectively. For comparison, we use primes $Q=5, 17$ and $67$,
resp., to achieve the corresponding transmission rates (slightly
higher though). From the figure, we see that augmenting the Golden
constellation with the CC  provides performance improvement of about
$1-1.5$ dB. The inferior performance of the augmented code for the
$2$ BPCU is due to the fact that the effective transmission rate
of the code, i.e., $\log_2(5)=2.32$, is significantly higher than
$2$ BPCU offered by Golden constellation. This choice is dictated by the
need to choose $Q$ as a prime so we have a nice lattice
representation for the received signal which is instrumental for
the MMSE-DFE Fano decoder.

\begin{figure}   
\begin{center}
\epsfig{file=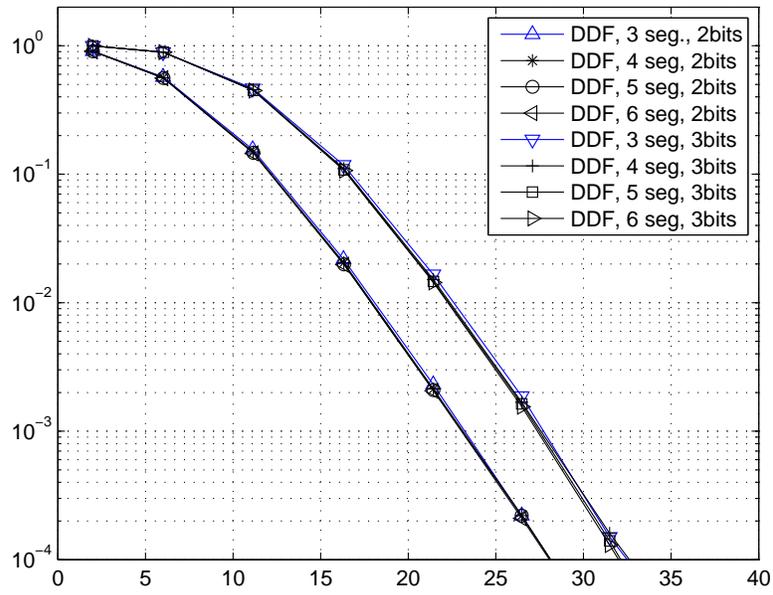,width=12.0cm}
\end{center}
\caption{Outage performance of DDF relay protocol with $3,4,5$ and $6$ segments.}
\label{ddf3456}
\end{figure}

\begin{figure}   
\begin{center}
\epsfig{file=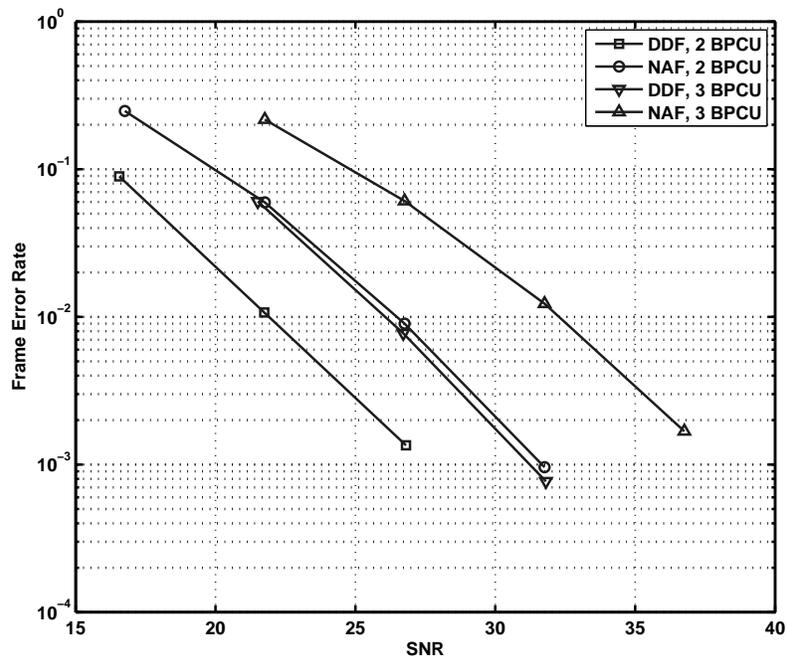,width=12.0cm}
\end{center}
\caption{Performance of DDF relay protocol (with 3 segments) {\it vs.} NAF relay protocol.}
\label{ddfvsnaf}
\end{figure}

Next, we proceed to the modified DDF protocol proposed in
Section~\ref{df}. At the source, the information stream is
appended with $16$ CRC bits, and the resulting vector is encoded
using a rate $1/4$ systematic CC. The relay attempts decoding
after waiting $N_i$ sub-blocks, where $N_i$ is the smallest among
the set of allowed waiting times such that the mutual information
at the relay exceeds the received rate. In order to avoid error
propagation at the destination, the decoded vector is checked for
validity using the CRC bits. If the decoder vector is valid,
i.e.,it satisfies the CRC check, then the relay uses the modified
DDF protocol. If the decoded stream at the relay does not satisfy
the CRC check, then the relay attempts decoding again at the next
allowed waiting time ($N_{i+1}$), and so on. At the receiver we
use the MMSE-DFE Fano decoder (note that the MMSE-DFE part is now
a trivial scaling since the channel seen by the destination is
SISO). We also assume that the destination knows the time at which
the relay starts transmitting (via overhead bits). For the range of
transmission rates considered in the sequel, it turns out that increasing
the number of segments beyond $3$ provides negligible increase in
performance. Figure~\ref{ddf3456} shows the outage probability of
DDF relay protocol, when the codeword partitioned into $3,4,5$ and $6$
segments. As seen from the figure, the gap between the outage curves is
negligible. Therefore, in the sequel, we consider only the variant of DDF
relay protocol with $3$ segments. Moreover, we choose the waiting fractions
$f_i$ according to Lemma~\ref{thrm:3}, {\it i.e.,} the pareto-optimal set of
waiting fractions for $3$ segments, given by $\{\frac{1}{2},\frac{2}{3}\}$.
Figure~\ref{ddfvsnaf} compares the performance of our variant of the DDF protocol
with that of the NAF protocol for $2$ and $3$ BPCU. The proposed DDF
strategy offers a gain of about $4$ dB( and about $6$ dB) over the NAF scheme,
for $2$( and $3$) BPCU. Note, however, that the DDF protocol
entails an complexity at the relay, as compared with the NAF
protocol, since the relay needs to decode the information stream. Finally,
we note that although several implementations of Amplify-and-Forward
and Decode-and-Forward variants exist
(\cite{stefanov1},\cite{duman1},\cite{nosratinia1}), all these works consider
low transmission rates. However, the impact of superior DMTs of our
schemes does lead to significantly better performance at low
transmission rates. It should be noted, however, that the difference in
performance between the protocols that are spectrally efficient and those
that are not manifests itself only at high transmission rate scenarios.
Therefore, we focus on high transmission rates in this paper.

\section{The Cooperative Multiple Access (CMA) Channel}\label{sec:cma}
In this section, we implement the CMA-NAF protocol for two users. We
start with describing the protocol. In the CMA-NAF protocol, each of
the two sources transmits once per cooperation frame, where a
cooperation frame is defined by two consecutive symbol-intervals.
Each source, when active, transmits a linear combination of the
symbol it intends to send and the (noisy) signal it received from
its partner during the last symbol-interval. For source $j$ and
frame $k$, we denote the broadcast and repetition gains by $a_j$ and
$b_j$, respectively, the symbol to be send by $x_{j,k}$, and the
transmitted signal by $t_{j,k}$. At startup the transmitted signals
will take the form
\begin{align}
t_{1,1} &= a_1x_{1,1} \label{eq:21} \\
t_{2,1} &= a_2x_{2,1}+b_2(ht_{1,1}+w_{2,1}) \label{eq:22} \\
t_{1,2} &= a_1x_{1,2}+b_1(ht_{2,1}+w_{1,1}) \label{eq:23} \\
t_{2,2} &= a_2x_{2,2}+b_2(ht_{1,2}+w_{2,2}) \label{eq:24}
\end{align}
where $h$ denotes the inter-source channel gain and $w_{j,k}$ the
noise observed by source $j$ during the frame $k$. (We assume that
$w_{j,k}$ has variance $\sigma_w^2$.) The corresponding signals
received by the destination are
\begin{align}
y_{1,1} &= g_1t_{1,1}+v_{1,1} \label{eq:25} \\
y_{2,1} &= g_2t_{2,1}+v_{2,1} \label{eq:26} \\
y_{1,2} &= g_1t_{1,2}+v_{1,2} \label{eq:27} \\
y_{2,2} &= g_2t_{2,2}+v_{2,2} \label{eq:28}
\end{align}
where $g_j$ is the gain of the channel connecting source $j$ to the
destination and $v_{j,k}$ the destination noise of variance
$\sigma_v^2$. Note that, as mandated by our half-duplex constraint,
no source transmits and receives simultaneously. The broadcast and
repetition gains $\{a_j,b_j\}$ are (experimentally) chosen to
minimize outage probability at the destination. As a consequence of
symmetry, $a_1$ and $a_2$, as well as $b_1$ and $b_2$, will have the
same optimal value. Thus, we assume that broadcast and repetition
gains are the same at each source and omit the subscripts, yielding
$\{a,b\}$. We also define a codeword as $N$ consecutive
symbol-intervals (assuming that $N$ is even, this means that there
are $N/2$ cooperation frames in each codeword). The following
result from \cite{AES:05} gives the diversity-multiplexing tradeoff
achieved by this protocol.
\begin{theorem}
The CMA-NAF protocol achieves the optimal diversity-multiplexing
tradeoff of the channel, i.e., for two users
\begin{align}
d(r) &= 2(1-r). \label{eq:29}
\end{align}
\end{theorem}

From (\ref{eq:21}-\ref{eq:28}), the received vector at the destination can be written as
\beq \label{cma_sysC}
\yv_1=\Hm_1\xv_1+\Bm\wv+\vv,
\enq
where $\yv=[y_{2,N}\ y_{1,N}\ \ldots\ y_{2,1}\ y_{1,1}]^{T}\in\CC^{N}$ is
the vector observed at the destination,
$\xv_1=[x_{2,N}\ x_{1,N}\ \ldots\ x_{2,1}\ x_{1,1}]^{T}\in\CC^{N}$ is the
vector formed by multiplexing the codewords transmitted by the two sources,
$\wv\in\CC^{N-1}$ and $\vv\in\CC^{N}$ are the AWGN vectors at the sources and
the destination, respectively. The effective channel matrix, $\Hm_1\in
\CC^{N\times N}$ is the upper triangular matrix
\be \label{cma_channel}
\Hm_1=a \Dm_{g}\left[\begin{array}{cccc} 1 & (bh) &
    \cdots & (bh)^N \\  & 1 &  \cdots & (bh)^{N-1} \\  & &
    \ddots &\vdots  \\  & & & 1\end{array}\right]
\ee
where $\Dm_{g}=\diag(g_2,g_1,g_2,...,g_2,g_1)$. Similarly, $\Bm\in \CC^{N\times(N-1)}$
is given by
\be \label{cma_noise}
\Bm=b \Dm_{g} \left[\begin{array}{cccc} 1 & (bh) &
    \cdots & (bh)^{N-1} \\  & 1 &  \cdots & (bh)^{N-2} \\  & &
    \ddots &\vdots  \\  & & & 1\\  &  & & 0 \end{array}\right]
\ee
The noise $(\Bm\wv+\vv)$ in (\ref{cma_sysC}) is colored, and must be whitened before tree
search with the MMSE-DFE Fano decoder can be performed.
Let $\Sigma=\sigma_w^2\Bm\Bm^{H}+\sigma_v^2\Id$ be the covariance matrix of $(\Bm\wv+\vv)$. Then,
\beq \label{cma_whiten}
\yv^c=\Hm^c\xv_1^c+\Sigma^{-\frac{1}{2}}(\Bm\wv+\vv)
\enq
where $\yv^c=\Sigma^{-\frac{1}{2}}\yv_1$, $\Hm^c=\Sigma^{-\frac{1}{2}}\Hm_1$
and the noise vector $\zv^c=\Sigma^{-\frac{1}{2}}(\Bm\wv+\vv)$ now consists of {\it i.i.d}
Gaussian components with variance $1$. $\Sigma^{-\frac{1}{2}}$ may be computed by any of the standard
methods, {\it i.e.,} singular value decomposition, Cholesky decomposition
or QR decomposition. The input-output relation in (\ref{cma_whiten}) can
now be written in the standard form (\ref{sys0}) by separating the real and
imaginary parts, to obtain the real $2N\times 2N$ system:
\beq
\yv=\Hm\xv+\zv
\enq
We construct the generator matrix $\Gm$ of the two sources
combined, as seen by the decoder, by appropriately multiplexing the rows and
columns of the lattice generators of the two sources, $\Gm_1$ and
$\Gm_2$. Fano decoding can now be done over the resulting joint channel-code
lattice, after MMSE-DFE preprocessing of the channel matrix $\Hm$. Finally,
the decoded codewords of the two sources are obtained by
demultiplexing the decoded lattice point.

\subsection{Numerical Results}

\begin{figure}   
\begin{center}
\epsfig{file=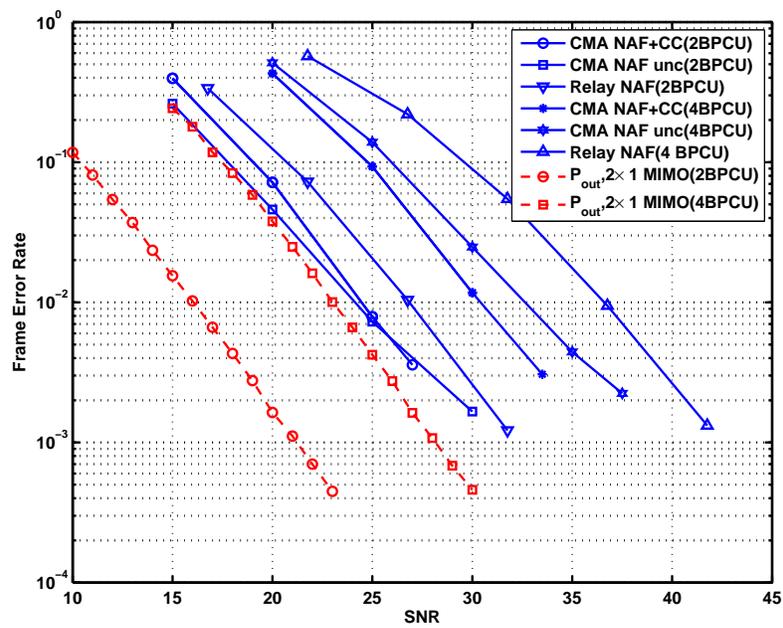,width=12.0cm}
\end{center}
\caption{FER performance of CMA-NAF {\it vs.} Relay NAF protocol.}
\label{cma_fer}
\end{figure}

\begin{figure}   
\begin{center}
\epsfig{file=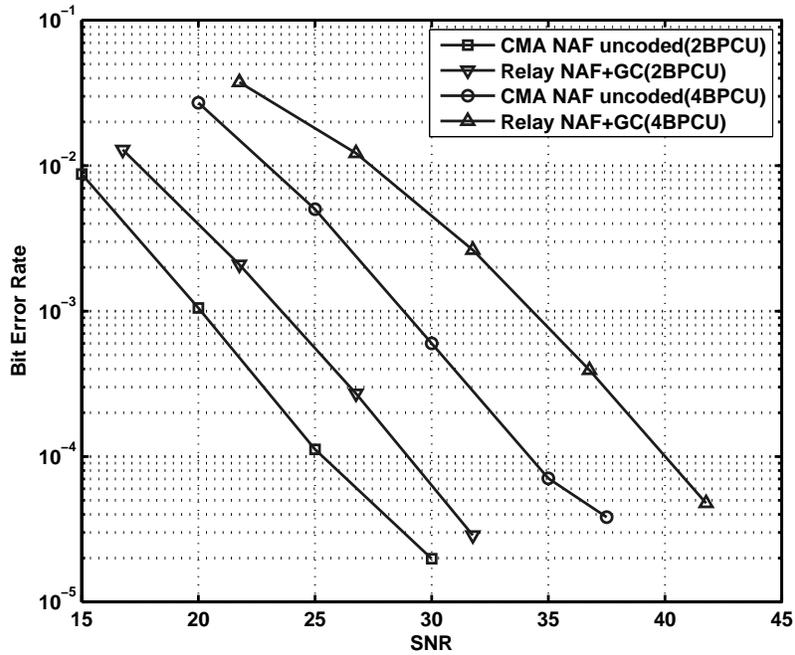,width=12.0cm}
\end{center}
\caption{BER performance of CMA-NAF protocol with uncoded QAM {\it vs.}
  Relay NAF protocol with Golden constellation.}
\label{cma_ber}
\end{figure}

\begin{figure}   
\centerline{\subfigure{\epsfig{file=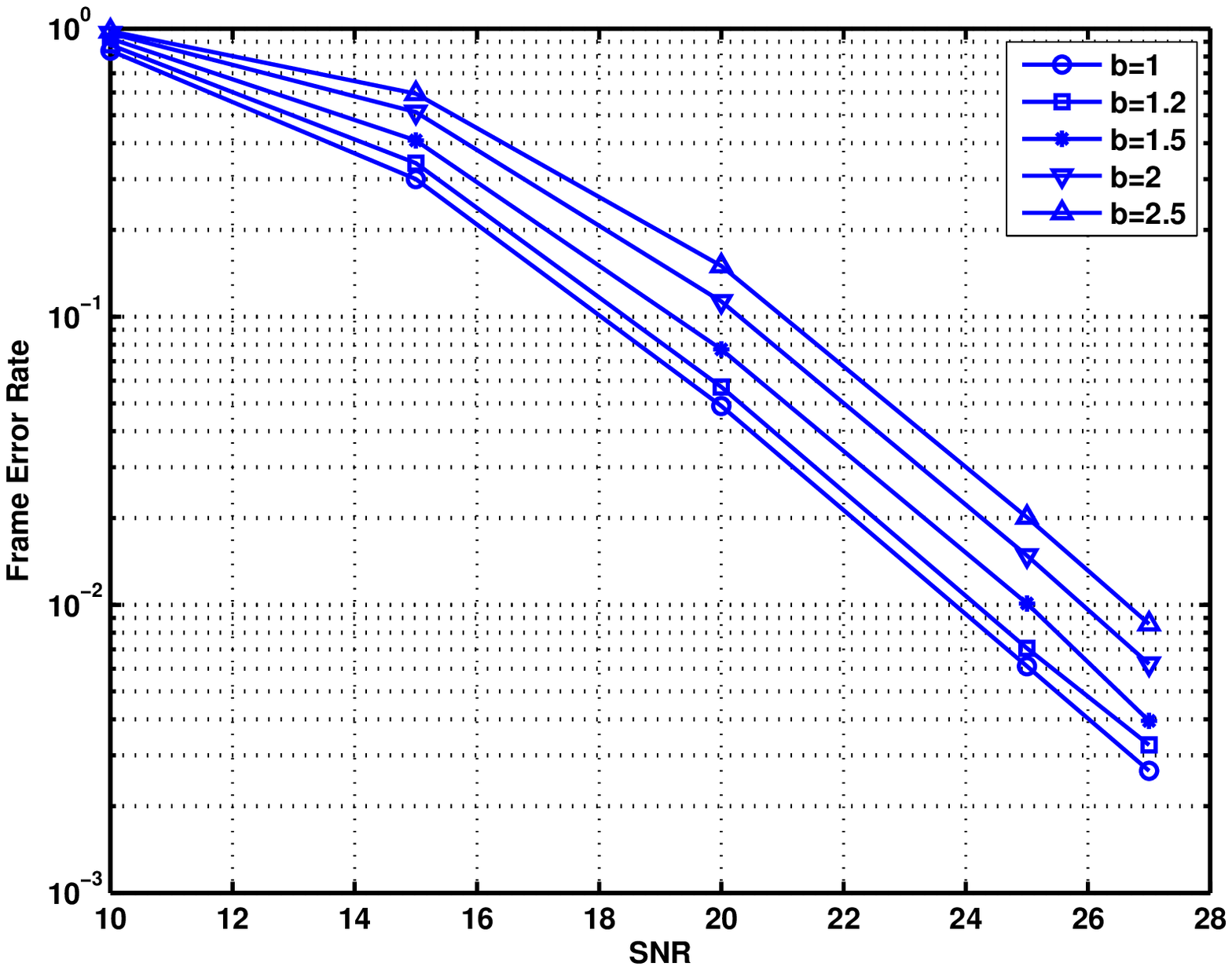,width=10.0cm}}
\hfill \subfigure{\epsfig{file=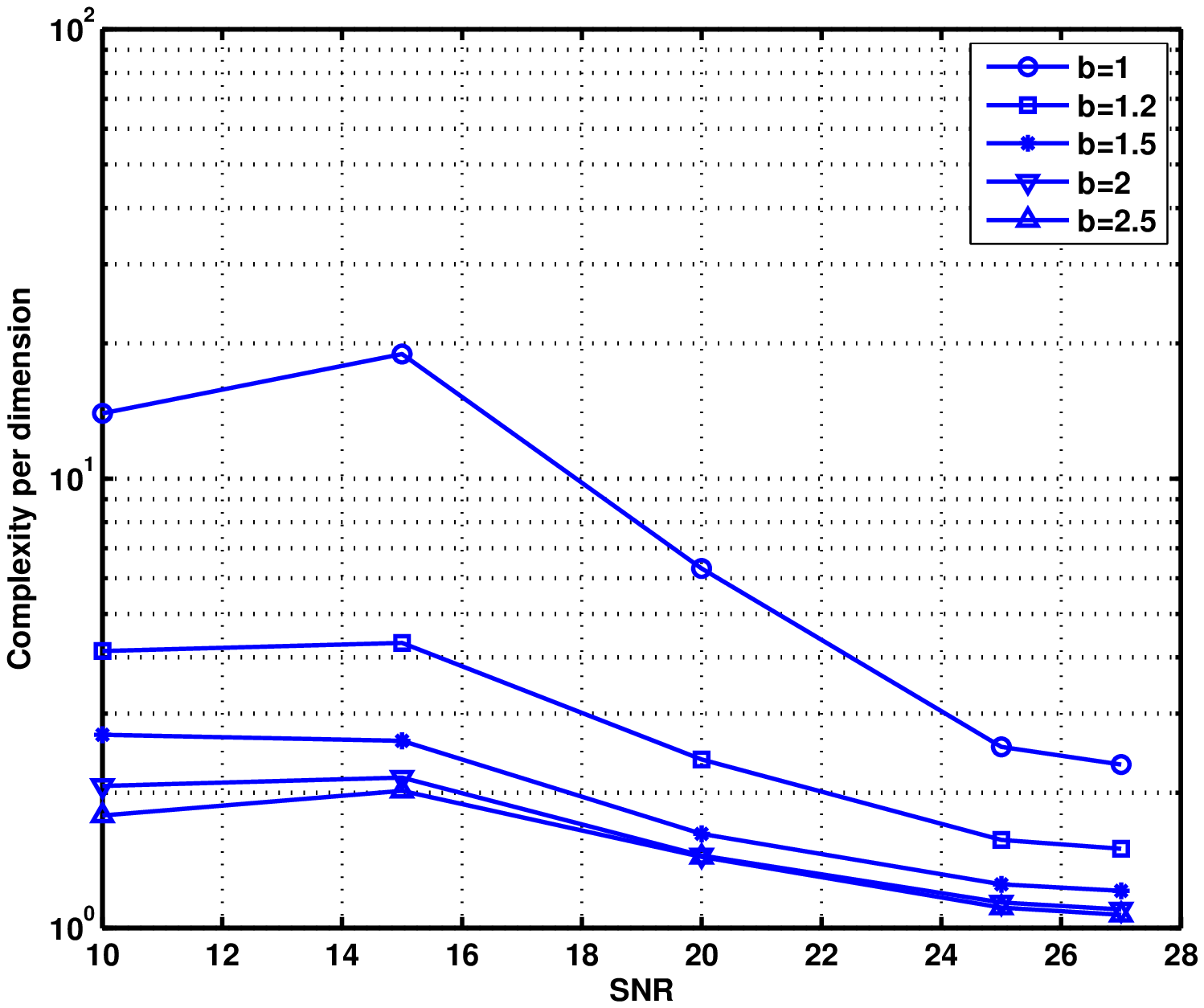,width=10.0cm}}}
\caption{Performance and Complexity of MMSE-DFE Fano decoder with
CMA NAF
  protocol, for $R=2$ BPCU.}
\label{cmabiasR2}
\end{figure}

\begin{figure}   
\centerline{\subfigure{\epsfig{file=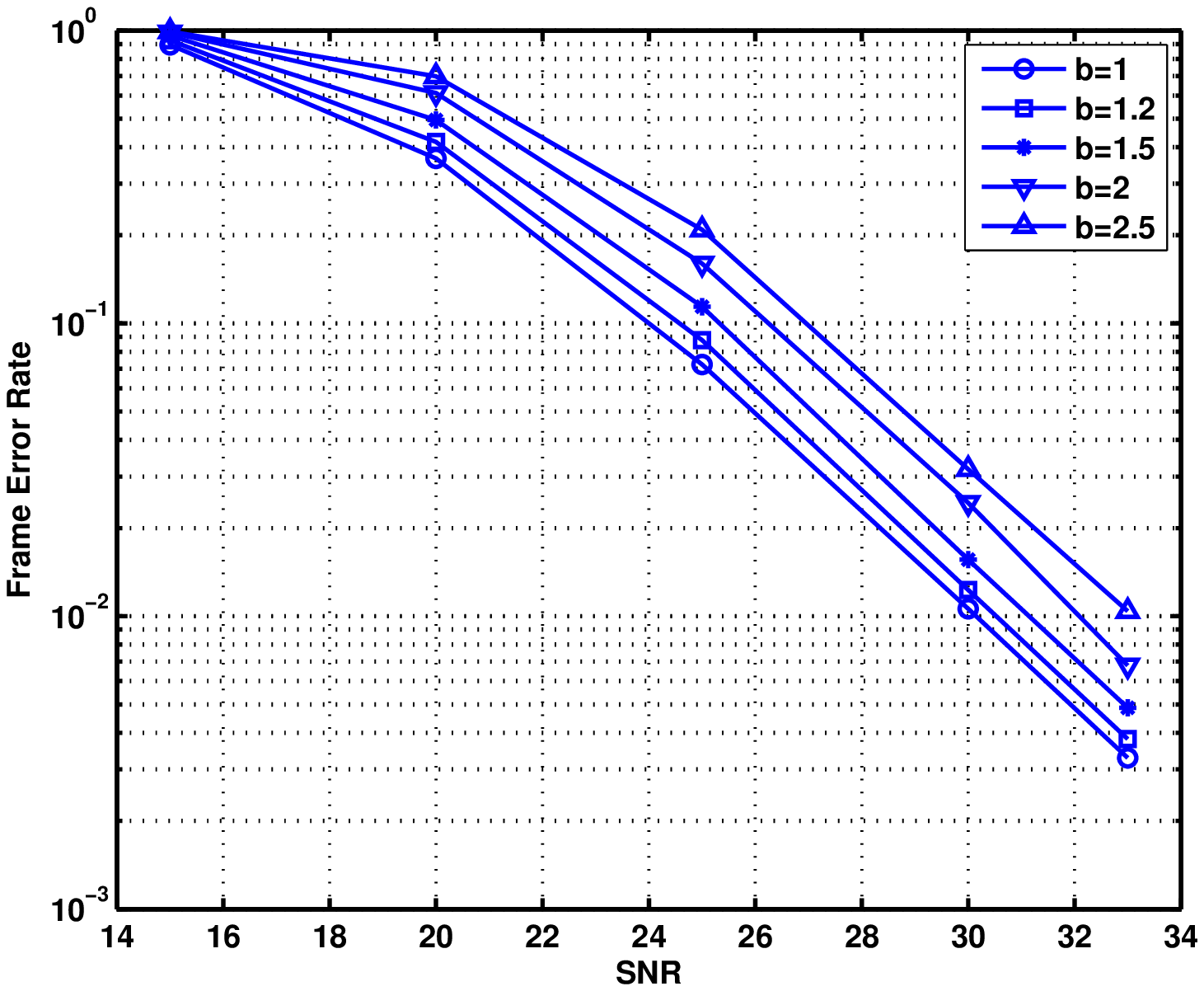,width=10.0cm}}
\hfill \subfigure{\epsfig{file=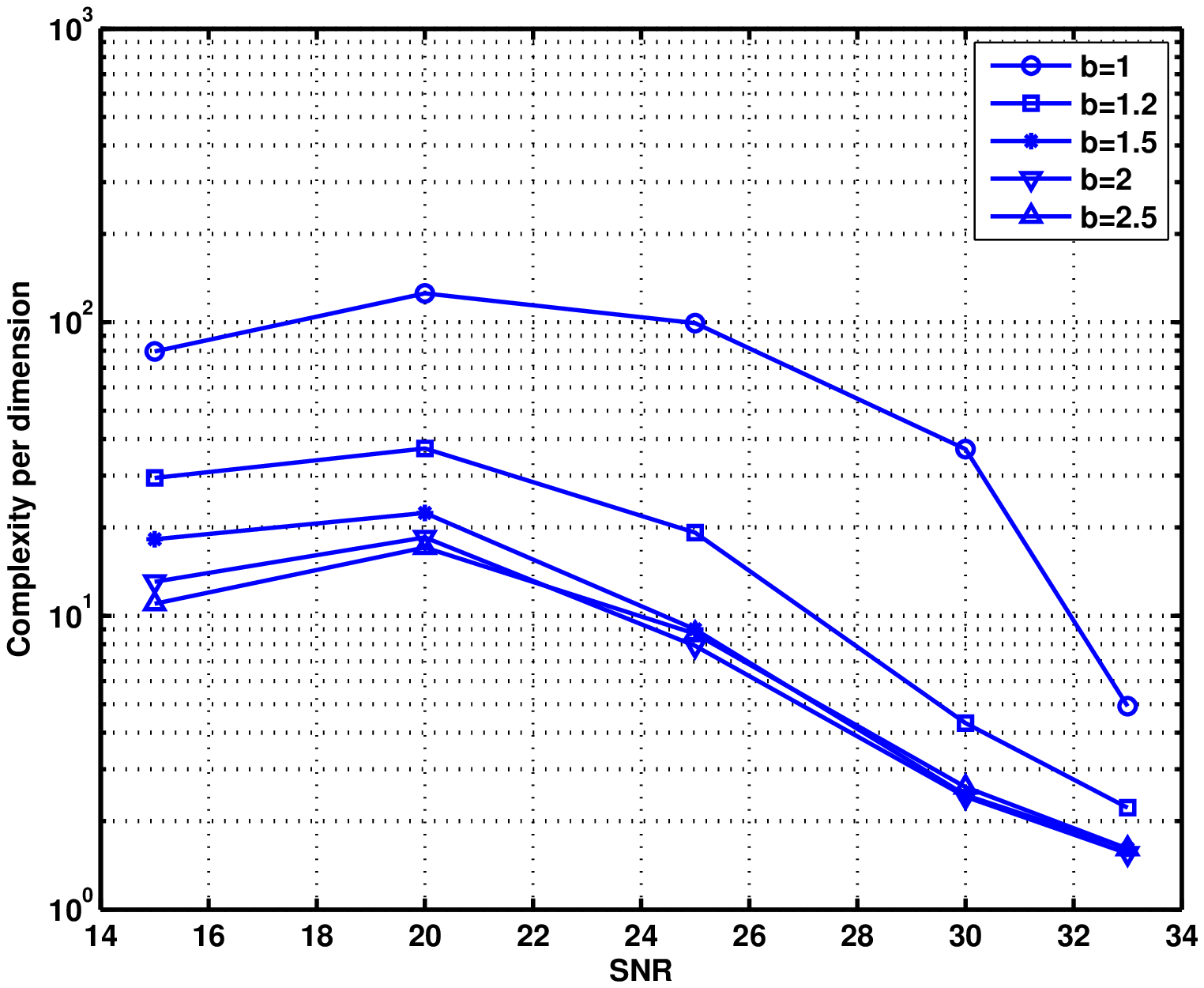,width=10.0cm}}}
\caption{Performance and Complexity of MMSE-DFE Fano decoder with
CMA NAF
  protocol, for $R=4$ BPCU.}
\label{cmabiasR4}
\end{figure}

In this section, we compare the performance of the lattice coded CMA-NAF
scheme with that of other schemes. The frame length is chosen to be
$N=128$. Each user encodes its information stream using a construction-A
lattice code obtained from a systematic convolutional code over $\ZZ_Q$.
At the destination, MMSE-DFE Fano decoding is performed over the joint
code-channel lattice of both the users. We define the frame error event as
$\{(\hat{\xv}_1,\hat{\xv}_2)\neq (\xv_1,\xv_2)\}$, where $\hat{\xv}_1$ and
$\hat{\xv}_2$ refer to the decoded codewords. Figure~\ref{cma_fer} shows
the frame error rate of the lattice coded CMA-NAF protocol {\it vs.} when
the two sources cooperate according to NAF relay protocol. We show the
performance for $2$ and $4$ BPCU, and the coding scheme achieving the
better performance (Golden constellation alone/Golden constellation+CC) is chosen for the NAF relay
protocol. Figure~\ref{cma_fer} also shows the frame error rate of the CMA-NAF
protocol, when both sources use uncoded QAM constellations for
transmission. As in Figure~\ref{nafrelay_eqpower}, the worse performance of
lattice coded CMA-NAF {\it w.r.t} uncoded QAM can be explained by the
higher transmission rate of the lattice code (with $Q=5$). Both coded and
uncoded transmission with CMA-NAF protocol perform significantly better
than the NAF-relay protocol. The performance gap between the two schemes
widens as the transmission rate increases, which can be explained by the
superior DMT of the CMA-NAF protocol over the NAF-relay protocol.
Figure~\ref{cma_ber} shows the bit error rate performance of
uncoded QAM transmission with CMA-NAF protocol, {\it vs.} NAF-relay
protocol with the Golden constellation. Again, CMA-NAF protocol
with uncoded transmission shows bit error rate improvement of about $3$ dB
at $2$ BPCU, and about $5$ dB at $4$ BPCU. Finally, Figures~\ref{cmabiasR2}
and \ref{cmabiasR4} show the performance and complexity trade-off of the MMSE-DFE Fano decoder {\it
  w.r.t} bias for the CMA NAF protocol with CC lattice coded
transmission (for $N=64$). We see from the figures that complexity of the
Fano decoder can be reduced at the expense of complexity. In our
simulations, we use $b=1.2$ to achieve good performance with reasonable complexity.

\section{Conclusions}\label{sec:conc}
We have developed a lattice theoretic framework for cooperative
coding in half-duplex outage-limited channels. This framework
achieves the cooperative diversity gains promised in \cite{AES:05}
and enjoys realizable encoding and decoding complexity. In the
relay channel, the proposed scheme achieves the optimal
diversity-multiplexing tradeoff for $r\leq 0.5$ while reducing the
channel between the source and relay, on one side, and the
destination, on the other side, to a time selective SISO channel.
This simplification allows for using traditional receiver
architectures in this scenario. A tree coding approach that
achieves the optimal diversity-multiplexing for the cooperative
multiple access channels is developed. In this context, the
MMSE-DFE Fano decoder is shown to yield an excellent
performance-vs-complexity tradeoff. The significant performance
gains offered by the proposed schemes are validated via a
comprehensive simulation study.

\bibliographystyle{ieeetr}
\bibliography{refr}

\newpage

\end{document}